\documentstyle[12pt,psfig]{article}
\begin{document}
\begin{titlepage}
\vspace{2.5cm}
\baselineskip 24pt
\begin{center}
\large\bf{ Covariant Wave Function Reduction}\\
\vspace{0.5cm}
\large\bf{ and Coherent Decays of Kaon Pairs$^1$}\\
\vspace{1.5cm}
\large{
Bernd A. Berg$^{2,3}$}\\
\vspace{0.5cm}
{berg@hep.fsu.edu}\\
\end{center}
\vspace{3cm}
\begin{center}
{\bf Abstract}\\
\end{center}

The recently developed relativistically covariant formulation of 
wave function reduction is illustrated for Lipkin's proposal to 
study CP violation in the coherent decay of kaon pairs. 
Covariant results are obtained in agreement with an amplitude 
approach proposed in the literature.


\vfill
\footnotetext[1]{{Work partially supported by the Department of 
Energy under contracts DE-FG05-87ER40319 and DE-FG05-85ER2500.}}
\footnotetext[2]{{Department of Physics, The Florida State University,
                      Tallahassee, FL~32306, USA. }}
\footnotetext[3]{{Supercomputer Computations Research Institute,
                      Tallahassee, FL~32306, USA.}}
\end{titlepage}

\baselineskip 24pt


About thirty years ago Lipkin~\cite{Li68} wrote an interesting paper
in which he proposed to exploit quantum mechanical correlations of the
EPR~\cite{EPR} type for experimental studies of CP violation in the
coherent decay of kaon pairs. His basic idea is quite simple: Consider
proton-antiproton annihilation into two neutral kaons $K^0K^0$, where
the intial state is assumed to be in an $s$-wave. First, disregard
$CP$ violation and take the states of definite $CP$, called $K_1$ and
$K_2$, as
basis for the $K^0$ meson system. Only the $K_1K_2$ final state is
allowed, whereas the states $K_1K_1$ and $K_2K_2$ are forbidden. This
result follows from parity conservation and Bose statistics. The initial
state has odd parity, while Bose statistics forbids odd-parity states
for two identical spinless bosons. The $K_1$ decays into two pions,
$\pi^0\pi^0$ or $\pi^+\pi^-$, whereas the $K_2$ does not. Therefore, we
have derived that only one of the $K^0K^0$ kaons can decay into two pions.
\medskip

With $CP$ violation the short- and long-lived kaons, $K_S$ and $K_L$,
form a natural (not necessarily orthogonal) basis of the neutral kaon
system and are both observed to decay into pions. Lipkin's observation
is that the above derivation requires only parity conservation and Bose
statistics, but does not depend on $CP$ conservation. $P\overline{P}$
annihilation creates the entangled intial state
\begin{equation} \label{entangled}
|i\rangle = {1\over \sqrt{2}}\, \left[ |K_L(x)\rangle\,|K_S(-x)\rangle
- |K_L(-x)\rangle\, |K_S(x)\rangle \right]
\end{equation}
where the $x$-axis is chosen as the direction of the momenta of the
kaons in their CM system; following Lipkin's notation
$(x)$ means that the particle is moving
in the positive $x$ direction and $(-x)$ means that it is moving in
the negative $x$ direction. Measurement of the decay of one of the kaons
into, say, $\pi^+\pi^-$ at a spacetime point
$(t_1,x_1)$~\cite{uncertainty} with $x_1>0$ forces the other into a
definite coherent mixture
\begin{equation} \label{K0r}
|K^0_r\rangle = \alpha\,|K_S(-x)\rangle + \beta\,|K_L(-x)\rangle 
\end{equation}
with coefficients $\alpha$ and $\beta$ such that at some time $t^0_2$
the matrix elements of
the two terms to the state $|\pi^+\pi^-\rangle$ cancel one another.
Further,
the subscript $r$ stands for ``reduced''. States of the type (\ref{K0r})
are then initial states for a beam of coherent neutral kaons. Similar
considerations hold for other processes such as $\Phi$-meson decay into
neutral kaons, or $\Upsilon (4S) \to B^0 B^0$~\cite{Ca81}.
\medskip

The initial $|K^0_r\rangle$ state (\ref{K0r}) is defined at the 
spacetime point $(t_r,x_r)$ where the reduction of the second kaon 
happens. The time $t_r$ is not necessarily identical with the time
$t^0_2$ at which the decay of $|K^0_r\rangle$ into 
$|\pi^+ \pi^-\rangle$ is forbidden.
Lipkin assumes that the spacetime point $(t_r,x_r)$ is
located on the instantaneous plane
\begin{equation} \label{instantaneous}
t_r=t_1\, .
\end{equation}
This reducion prescription lacks relativistic covariance (see below)
and a few years ago
Kayser and Stodolsky~\cite{Ka95} proposed as a remedy to abolish the
concept of wave function reduction in favor of an amplitude approach
within which they obtained covariant results. Recently, the author
succeeded~\cite{Be98} in formulating a relativistic version of the
reduction postulate. In this paper I shortly summarize the basic ideas
of covariant reduction and derive the relativistic equations for some
of Lipkin's results. Agreement is found with ref.\cite{Ka95}.
\medskip

Figure~1 depicts the instantaneous reduction scenario
(\ref{instantaneous}) in the CM frame of the kaons. The $K^0K^0$ pair
is created at the origin $0$ of the depicted reference frame and each
kaon propagates on a straight line within the forward lightcone of
$0$. The symmetry of the branches under $x\to -x$ comes from the
almost degenerate kaon masses. It is assumed
that pion measurements allow to reconstruct the decay
spacetime point $(t_1,x_1)$ as indicated, and reduction at $(t_r,x_r)$
is implied in the other branch. Decay of the second kaon may be
observed at some later time $(t_2,x_2)$.
Now, instantaneous reduction faces
difficulties with relativistic covariance. For instance, the LAB frame
of the detectors can be distinct from the CM frame of the kaons. Then
the LAB frame times $t'_r$ and $t'_1$ (corresponding to the CM
spacetime points $(t_r,x_r)$ and $(t_1,x_1)$ transformed into the
LAB frame) will no longer be instantaneous.
Further, in case that $(t_1,x_1)$ and $(t_2,x_2)$ are spacelike, the 
time ordering $t_2>t_r$ of the CM frame can change into $t'_2<t'_r$ in
the LAB frame. This is shown in figure~1 where the upper dotted line
originating at $(t_1,x_2)$ indicates the instantaneous time plane
of a LAB frame, where the relative velocity between
the CM and LAB frame is assumed to be $0.447\,c$ along the
$x$ direction (no care is taken to match the kinematic conditions of
proton and kaons).
Nothing is gained by embarking on the business of preferred reference
frames. It is by no means obvious whether the rest frame of the kaons
or the LAB frame of the detectors (after all they cause the reduction?)
should be preferred. Desireable is a covariant generalization of the
reduction process, which emerged only recently~\cite{Be98}.
\medskip


Figure~2 depicts the situation of figure~1 from the viewpoint of
measurement at the spacetime point $(t_1,x_1)$. The wave function can
be interpreted as carrier of information, namely it allows to extract
probabilities for the results of eventual measurements. At the
moment when a measurement has been carried out and produced a
definite result, the wave function may undergoe a discontinuous
jump due to incorporating the result. For a measurement at $(t_1,x_1)$
the information available comes from its past. This means from within 
or from the surface of the backward light cone (BLC) of this spacetime 
point. It follows that
the measurement resets the the wave function (as carrier of
information) on the BLC of $(t_1,x_1)$. This somewhat surprising 
result is at second thought rather obvious: (A)~Information of the past
({\it i.e.} the wave function inside the BLC) cannot be changed, because
that would change the probabiltities for the (already carried out)
measurement at $(t_1,x_1)$.
(B)~The newly gained information applies to the future of all points
of the region over which the information for the measurement was
gathered. For instance, by observing a supernova explosion 10,000
lightyears away, we obtain information about what is going on there
10,000 years ago and not about what is going on there now.
\medskip

A few more points deserve to be mentioned. (A)~In view of interference
effects the author favors to identify the wave function with physical
reality~\cite{Be98}. This does not exclude the interpretation as
carrier of information. Just, the information is reality too.
(B)~Light cone sections
are the only hypersurfaces which allow for a covariant formulation
of reduction. (C)~Consistency of multiple spacelike measurements leads
to some complications and a consistent formulation is given
in~\cite{Be98}. (D)~An earlier attempt~\cite{He70} to formulate
reduction on the BLC failed~\cite{Ah80,Be98} and ref.\cite{Ah80}
embarked on a
position similar to that of ref.\cite{Ka95}. In figure~2 the reduction
spacetime point for the second kaon is indicated as $(t_r,x_r)$. I
proceed to show that this gives a relativistically covariant
description of the reduction process.
\medskip

Before the decay, the paths of the two kaons are given by
$x_i = v_i\, t_i$ where the $v_i$, $i=1,2$ are their velocities.
The reduction time $t_r$ follows from the equations
$x_r = v_2\, t_r$ and $x_r = x_1 + c\, (t_r - t_1)$. Elimination of
$x_r$ gives
\begin{equation} \label{blc1}
 t_r = {c-v_1 \over c - v_2}\ t_1\, .
\end{equation}
The non-relativistic limit $c\to\infty$ reproduces
equation~(\ref{instantaneous}) and in the
extreme relativistic limit $v_1\to c$ we get $t_r\to 0$.
Equation~(\ref{blc1}) does not look very covariant, but it
is. Using rapidities defined by $\tanh (\zeta_i) = v_i/c$, $i=1,2$
and proper times $\tau_1 = t_1/\cosh(\zeta_1)$,
$\tau_r = t_r/\cosh(\zeta_2)$ this is easily seen.
Equation~(\ref{blc1}) becomes
\begin{equation} \label{blc2}
\tau_r = R (\zeta_1,\zeta_2)\, \tau_1 ~~{\rm with}~~
R(\zeta_1,\zeta_2) = {\cosh(\zeta_1)-\sinh(\zeta_1)
\over \cosh(\zeta_2) - \sinh(\zeta_2)}\, .
\end{equation}
Let us tranlate this relation into an inertial frame that moves which
respect to the original one of figures~1 and~2 with rapidity $\eta$.
The proper times are scalars and stay invariant. Hence, the factor
$R(\zeta_1,\zeta_2)$
has to stay invariant too. The rapidities simply transform according
to $\zeta_i \to \zeta_i + \eta$, $i=1,2$ and the invariance of
$R(\zeta_1,\zeta_2)$ follows from the identity
$$ \cosh (\zeta + \eta) - \sinh (\zeta + \eta) =
\left( \cosh (\eta) - \sinh (\eta) \right)\,
\left( \cosh (\zeta) - \sinh (\zeta) \right)\, .$$
Imagine now that some decay mode $D_1$ is observed at $(t_1,x_1)$ in
the positive $x$ direction. Equation~(\ref{entangled}) implies 
at $(t_r,x_r)$ the reduced state 
\begin{equation} \label{rstate}
|K^0_r\rangle = e^{-i\,(m_L\,\tau_1 - m_S\,\tau_r)}\,
\langle D_1|T|K_L\rangle\, |K_S(-x)\rangle -
e^{-i\,(m_L\,\tau_r - m_S\,\tau_1)}\,
\langle D_1|T|K_S\rangle\, |K_L(-x)\rangle 
\end{equation}
where the masses $m_L$ amd $m_S$
are complex to deal with the finite kaon lifetimes. The proper
times in this equation are related by (\ref{blc2}) and from this
it is obvious that the reduction is Lorentz covariant. At some later
time $t_2$, corresponding to $\tau_2$ for the proper time of the
second kaon, a decay mode $D_2$ may be observed in negative $x$
direction. For $D_2=D_1$ it follows from (\ref{rstate}) that the
probability for that decay vanishes when the proper times
agree, {\it i.e.} at $\tau_2=\tau_1$. This result is in agreement
with ref.\cite{Ka95} and the same holds
for other physical quantities.
\medskip

Some readers may wonder about the possibility of reversing the
order of reductions in case that $(t_1,x_1)$
and $(t_2,x_2)$ are spacelike. Because spacelike operators
commute both orders of BLC reductions are allowed and each gives 
a well-defined wavefunction at all times, see~\cite{Be98} for
details. This solves an old problem~\cite{Ah80}:
In non-relativistic quantum theory complete sets of commuting
operators exist which uniquely determine the wave function. Does
this still hold in relativistic quantum field  theory? The answer
is no by explicit construction: For $n$ spacelike measurements at
least $n!$ distinct wave functions exist which are consistent with
these measurements.
\medskip

Other reduction prescriptions do also allow to calculate covariant
amplitudes. However, reduction on the backward lightcone stands out
as the unique possibility to mantain a well-defined space-time
picture of the wave function. This contributes to conceptional
clarity and computational simplicity. Finally,
numerous experiments found violations
of Bell inequalities~\cite{Bell}, see ref.\cite{Aspect}
for a few examples and further references. Therefore, reality 
as imagined by EPR~\cite{EPR} is
now ruled out on experimental grounds. Covariant wave function
reduction opens the possibility to take the wave function itself
as a reality substitute.
\bigskip

\noindent
{\bf Acknowledgement:} I would like to thank Leo Stodolsky for drawing
my attention to ref.\cite{Ka95} and Peter Weisz for his hospitality when
I visited the Munich Max Planck Institut f\"ur Physik und Astrophysik.


\bigskip

\begin{thebibliography}{12}

\bibitem{Li68} H.J. Lipkin, Phys. Rev. {\bf 176}, 1715 (1968).

\bibitem{EPR} A. Einstein, B. Podolsky and N. Rosen, Phys. Rev.
              {\bf 47}, 777 (1935).

\bibitem{uncertainty} In connection with the velocities (later used
in this paper) uncertainty relations forbid precise measurement of
all quantities introduced. To impose such limited accuracy is
consistent with the other considerations of this paper. 

\bibitem{Ca81} A.B. Carter and A.I. Sanda, Phys. Rev. {\bf D23},
1567 (1981); I.I. Bigi and A.I. Sanda, Nucl. Phys. {\bf B193},
85 (1981).

\bibitem{Ka95} B. Kayser and L. Stodolsky, Phys. Lett. {\bf B359},
343 (1995).

\bibitem{Be98} B. Berg, quant-ph/9807046, submitted to Phys. Rev. A.

\bibitem{He70} K.-E. Hellwig and K. Kraus,
Phys. Rev. {\bf D1}, 566 (1970).

\bibitem{Ah80} Y. Aharanov and D.Z. Albert,
Phys. Rev. {\bf D21}, 3316 (1980).

\bibitem{Bell} J.S. Bell, Physica {\bf 1}, 195 (1964); J.D. Franson,
Phys. Rev. Lett. {\bf 62}, 2205 (1989).

\bibitem{Aspect} A. Aspect, J. Dalibard, and 
G. Roger, Phys. Rev. Lett. {\bf 49}, 1804 (1982);
P.R. Tapster, J.G. Rarity and P.C.M. Owens,
Phys. Rev. Lett. {\bf 73}, 1923 (1994);
W. Tittel, J. Brendel, B.Gisin, T. Herzog, H. Zbinden and N. Gisin,
Phys. Rev. {\bf A57}, 3229 (1998).
\end{thebibliography}
\end{document}